\documentclass[fleqn,usenatbib]{mnras}

\usepackage{newtxtext,newtxmath}

\usepackage[T1]{fontenc}
\usepackage{ae,aecompl}

\usepackage{graphicx}	
\usepackage{amsmath}	
\usepackage{amssymb}	

\usepackage[UKenglish]{babel}
\usepackage{bigfoot}

\defcitealias{Kegerreis+2018}{K18}

\title[Planetary Giant Impact Simulations]{
  Planetary Giant Impacts:
  Convergence of High-Resolution Simulations
  using Efficient Spherical Initial Conditions and SWIFT
  }

\author[J. A. Kegerreis et al.]{J. A. Kegerreis$^{1}$\thanks{
  \href{mailto:jacob.kegerreis@durham.ac.uk}{jacob.kegerreis@durham.ac.uk}},
  V. R. Eke$^{1}$,
  P. Gonnet$^{2}$,
  D. G. Korycansky$^{3}$,
  R. J. Massey$^{1}$,
  \newauthor
  M. Schaller$^{4}$,
  L. F. A. Teodoro$^{5}$
  \\
  $^{1}$Institute for Computational Cosmology, Durham University, Durham, DH1 3LE, UK\\
  $^{2}$Google AI Perception, Google Switzerland, 8002 Zurich, Switzerland\\
  $^{3}$CODEP, Earth Sciences, University of California, Santa Cruz, CA, USA\\
  $^{4}$Leiden Observatory, Niels Bohrweg 2, 2333 CA Leiden, Netherlands\\
  $^{5}$BAERI/NASA Ames Research Center, Moffett Field, CA, USA
  }

\date{Accepted XXX. Received YYY; in original form ZZZ}

\pubyear{2019}

\begin{document}
\label{firstpage}
\pagerange{\pageref{firstpage}--\pageref{lastpage}}
\maketitle

\begin{abstract}
We perform simulations of giant impacts onto the young Uranus using smoothed particle hydrodynamics (SPH) with over 100 million particles. This $100$--$1000\times$ improvement in particle number reveals that simulations with below $10^7$ particles fail to converge on even bulk properties like the post-impact rotation period, or on the detailed erosion of the atmosphere. Higher resolutions appear to determine these large-scale results reliably, but even $10^8$ particles may not be sufficient to study the detailed composition of the debris -- finding that almost an order of magnitude more rock is ejected beyond the Roche radius than with $10^5$ particles. We present two software developments that enable this increase in the feasible number of particles. First, we present an algorithm to place any number of particles in a spherical shell such that they all have an SPH density within 1\% of the desired value. Particles in model planets built from these nested shells have a root-mean-squared velocity below 1\% of the escape speed, which avoids the need for long precursor simulations to produce relaxed initial conditions. Second, we develop the hydrodynamics code SWIFT for planetary simulations. SWIFT uses task-based parallelism and other modern algorithmic approaches to take full advantage of contemporary supercomputer architectures. Both the particle placement code and SWIFT are publicly released. 
\end{abstract}




\section{Introduction}
\label{sec:introduction}

Giant impacts are thought to dominate
many planets' late accretion and evolution.
We see the consequences of these violent events
on almost every planet in our solar system,
from the formation of Earth's Moon
to the odd obliquity of Uranus spinning on its side.
As such, they are expected to play a similarly important role
in the evolution of the many exoplanetary systems
that are now being observed in detail.
These complicated and highly non-linear processes
are most commonly studied using
smoothed particle hydrodynamics (SPH) simulations \citep[e.g.][]{Benz+1986}.

SPH is a Lagrangian (particle-based) method
used in a wide range of topics in astrophysics and many other fields,
from planetary impacts and supernovae
to galaxy evolution and cosmology
\citep{Springel2010,Monaghan2012}.
As well as correctly evolving the simulation particles with time,
it is crucial to start from appropriate initial conditions
for any model's evolution to accurately reflect its real-world counterpart.
Furthermore, enough particles must be used
to resolve the physical processes in sufficient detail,
and recent work has shown that standard-resolution simulations 
($10^5$ to $10^6$ particles)
can produce unreliable results that have not converged numerically
\citep{Hosono+2017,Genda+2015}.
This motivates the pursuit of simulation codes that can take full advantage of
contemporary supercomputing architectures,
enabling more particles to be used to run suitable convergence tests
and, hopefully, simulations with sufficiently high resolution.

Towards this end, we present
the simple SEA\footnote{
  The SEAGen code is publicly available at
  \href{https://github.com/jkeger/seagen}{github.com/jkeger/seagen}
  and the python module \verb|seagen| can be installed directly with
  \href{https://pypi.org/project/seagen/}{pip}.
  }
scheme for creating optimal spherical arrangements of particles
(\S\ref{sec:methods:particles}, \S\ref{sec:results:particles})
and the hydrodynamics code SWIFT\footnote{
  SWIFT is in open development and is publicly available at
  \href{http://swiftsim.com/}{www.swiftsim.com}.
  }
that we have developed to run planetary impact (and cosmological) simulations
(\S\ref{sec:methods:swift}).
We use these tools to model giant impacts onto a young Uranus
at high resolution using over $10^8$ SPH particles,
and test the convergence of various physical properties
with increasing particle number
(\S\ref{sec:intro:uranus}, \S\ref{sec:results:uranus}).
We then present conclusions in \S\ref{sec:conclusions}.

\subsection{Particle Placement and Initial Conditions}
\label{sec:intro:particles}

Many problems in astrophysics feature spherical symmetry,
such as those involving stars or planets.
Before one can simulate and study these problems
with a particle-based method like SPH,
each initial object must first be converted into
an appropriate set of particles.
Two common approaches to creating arrangements of SPH particles in spheres are:
(1) to use a lattice that can be distorted
until it approximately matches the required shape and densities;
and (2) to relax an imperfect initial state into a fully settled one
with a pre-production simulation.

A third, more recent approach is to arrange the particles analytically
while accounting for the spherical symmetry from the outset,
by placing particles in nested spherical shells
\citep{Saff+Kuijlaars1997,Raskin+Owen2016a,Reinhardt+Stadel2017}.
These methods aim to combine the minimal computation
required for lattice methods
with the settled and symmetric properties of simulated glasses.
We present a comparable scheme that further ensures
every particle's SPH density is within 1\% of the desired value.
This leads to initial conditions that are quick and simple to produce,
close to equilibrium, and in which every particle has a realistic density and,
therefore, pressure.

Lattice-based methods are popular because they are easy to implement
and, since the inter-particle separations are uniform by construction,
they can accurately match a simple density profile.
This can be achieved either by stretching the lattice radially
or by varying the particle mass --
although keeping the masses of all particles very similar
is usually desirable.
However, the grid-like properties of a lattice introduce
unwanted anisotropies to a problem
and may be unstable to perturbations
\citep{Herant1994,Morris1996,Lombardi+1999}.

Furthermore, a spherically symmetric object like a planet or star
features important boundaries at specific radii.
Both the outer surface and internal layers
require discontinuities in density and material.
The particles in a lattice are dispersed at all radii,
so cannot reproduce such sharp changes at these boundaries.
A similarly quick and simple alternative to lattice methods is to
randomly place particles following an
appropriate probability distribution function,
either restricted in nested shells or anywhere in the sphere.
However, these methods are noisy and result in
extreme variations in local particle densities.

In SPH, the density of a particle is estimated by
summing the masses of typically $\sim$50 nearby `neighbour' particles,
weighting by a 3D-Gaussian-shaped kernel
that decreases the contribution of more-distant neighbours.
Thus a particle that is placed too close to another
will have a higher density and not be in equilibrium.
The accuracy of every particle's density is important
because of how stiff the equation of state (EoS) can be for a material,
such as the granite planetary example we test here.
This means that a slightly too-dense particle will be assigned
a dramatically too-high pressure by the EoS,
leading to unphysical behaviour as soon as the simulation is started.
In the case of a tabulated EoS,
this may also cause practical problems by pushing a particle
outside of the parameter space covered by the tables.

An obvious improvement on these crude analytical distribution methods
is to run a simulation that iterates the initial particle positions
towards a more stable state.
One approach is to use an inverse gravitational field to
repel the particles from each other \citep{Wang+White2007}.
A more sophisticated version of this was developed by \citet{Diehl+2015}
based on weighted Voronoi tessellations.
Another method is to add a damping force to reduce any transient velocities
as the particles are allowed to evolve
under otherwise-normal gravitational and material pressure forces.
In all cases, the simulation is run until a condition is met
to call the system `relaxed',
such as when the particle velocities or accelerations reach some small value.

These methods can generate particle configurations that are stable and relaxed,
but at a cost of performing extra simulations.
Especially for large numbers of particles,
this can be a computationally expensive process
and can take large amounts of time,
comparable to the final simulation for which
the initial conditions are being generated.
Depending on the method used, the particles may also
settle to a distribution somewhat different to the desired initial profile.

The spherical symmetry and sharp radial boundaries of astrophysical objects
strongly motivate the arrangement of particles in nested spherical shells.
If the particles could be distributed uniformly in each shell,
then no computationally expensive simulation would be required
to create relaxed initial conditions.
However, the equidistant distribution of points on the surface of a sphere
is a challenging problem,
and has been studied for applications in a wide variety of fields:
from finding stable molecular structures like buckminsterfullerene
to making area-integral approximations,
in addition to the pure mathematical curiosity of
such a trivial question in 2D
(equally spaced points on a circle)
becoming so complicated in higher dimensions \citep{Saff+Kuijlaars1997}.

Similar ideas motivated the work of
\citet{Raskin+Owen2016a} and \cite{Reinhardt+Stadel2017},
who both presented algorithms for arranging particles in spherical shells.
One issue with \citet{Raskin+Owen2016a}'s method is that
in each shell there are a few particles with large overdensities,
placing the particles slightly out of equilibrium
(see \S\ref{sec:results:particles}).
\citet{Reinhardt+Stadel2017} divide the sphere into equal regions
that can be further subdivided (using the HEALPix scheme),
with the disadvantage that only
sparsely distributed numbers of particles 
($12 \times 4^n$ for $n \in \mathbb{N}$) 
can be placed in each shell.
Furthermore, some particles in each shell show SPH densities more than 5\%
discrepant from the desired profile density (their Fig.~4).

In \S\ref{sec:methods:particles},
we present an algorithm for arranging
any number of particles in a spherical shell
such that every particle has an SPH density within 1\% of the median.
Our method involves a simple division of the sphere into equal-area regions
arranged in latitudinal collars,
followed by slightly stretching the collars away from the poles.
Concentric shells can then be set up to precisely follow an
arbitrary radial density profile,
taking care to align the shells with any radial boundaries.
We apply this stretched equal-area (SEA) algorithm to create
near-equilibrium models of planets, 
and present the results in \S\ref{sec:results:particles}.

\subsection{Convergence and Uranus Giant Impacts}
\label{sec:intro:uranus}

The need to increase resolution to improve studies of existing topics 
was recently demonstrated by \citet{Hosono+2017}.
Concerningly, they found giant impact simulations that gave apparently
reliable results with up to $10^6$ particles
had not actually converged when re-tested with $10^7$--$10^8$.
\citet{Genda+2015} also found incomplete convergence 
of disruptive impact simulations with up to $5\times10^6$ particles.

A numerically converged result
is not necessarily physically correct.
For example, several studies 
\citep[e.g.][]{Woolfson2007,Deng+2019a}
have pointed out the difficulties for SPH in modelling 
the interaction of multiple materials
or the treatment of density discontinuities,
which may not be immediately fixed by higher resolutions.
That said, it is crucial that we at least obtain a reliable answer
to the (imperfect or not) question that we ask the computer to solve,
so convergence is an important first step.

As an example with which to investigate convergence 
and test the simulation tools presented in this paper,
we consider the giant impact that likely knocked over 
the planet Uranus to spin on its side.
Previously, we ran SPH simulations 
to study the consequences of this violent event
using $\sim$$10^6$ particles \citep[][hereafter \citetalias{Kegerreis+2018}]{Kegerreis+2018} --
as an improvement on the $<$$10^4$ particles
in the single previous study by \citet{Slattery+1992} almost 30 years ago.
As well as confirming that the impact can explain Uranus' spin,
we found that with a grazing collision
the impactor could form a thin shell around the planet's ice layer,
perhaps trapping the interior heat to help explain
the freezing outer temperatures.
$\sim$10\% of the target's atmosphere 
becomes unbound to escape from the system 
and a small amount of the impactor's rocky core
is ejected into the debris disk.
\citet{Kurosaki+Inutsuka2019} recently explored a different, complementary part
of the wide parameter space with $\sim$$10^5$ SPH particle simulations.
They varied the entropy of the proto-Uranus target
to examine the effects on the angular momentum and the debris.

In \S\ref{sec:methods:swift} we summarise the SWIFT hydrodynamics code
and its development to run these planetary simulations
and take advantage of contemporary supercomputer architectures.
In \S\ref{sec:results:uranus} we use SWIFT and the SEA particle placement method 
to repeat simulations of Uranus giant impacts
from \citetalias{Kegerreis+2018} using $10^5$ up to $10^8$ SPH particles,
and test the convergence of the post-impact planet's rotation rate,
the erosion of the atmosphere, 
and the ejection of rocky material into the debris disk.

\section{Methods}
\label{sec:methods}

\subsection{Particle Placement and Initial Conditions}
\label{sec:methods:particles}

The goal is to distribute a number of similar-mass particles in a sphere,
such that the SPH density of every particle
accurately matches a given density profile
(see \S\ref{sec:profiles}).
In order to follow an arbitrary radial profile
that may include sharp discontinuities,
such as a core-mantle boundary or a planet's surface,
it is convenient to distribute the particles in spherical shells.
The particles can then be assigned any property using other radial profiles,
such as their material type and temperature or internal energy.

The two inputs for this problem are the desired total number of particles
and the radial density profile.
The profile is first used to find the enclosed mass at each radius.
The number of particles then gives the nominal particle mass.
We iterate outwards from the centre, placing particles in successive shells,
following the density profile.
First, we must determine the radius of each shell and how many particles
are required to account for its mass (\S\ref{sec:method:shells}).
Then, the question is how to arrange
an arbitrary number of particles on a spherical shell,
for which we describe our stretched equal-area (SEA) method
(\S\ref{sec:method:particles}).

\subsubsection{Shells and Layers}
\label{sec:method:shells}

We begin by placing a tetrahedron of particles near the centre,
so the first `shell' is actually the sphere
that encloses the mass of four particles.
If this central sphere has radius $dr_c$ and density $\rho_c$,
then the thickness, $dr$, of a subsequent shell with density $\rho$ is
\begin{equation}
	dr = dr_c \left( \dfrac{\rho_c}{\rho} \right)^{1/3} \;.
	\label{eqn:dr_shell}
\end{equation}
The number of particles in a shell is then simply the mass of that shell
divided by the nominal particle mass.
This must be rounded to an integer,
giving an actual particle mass in each shell
that may be slightly different to the nominal mass.
This amounts to maximum deviations of $\sim$1\% for $10^6$ total particles 
and $\sim$0.1\% for $10^8$.
The shell thickness could be tweaked instead 
to enforce strictly equal particle masses.
The particles in the shell are then all assigned
the same properties (e.g. temperature),
set by the mass-weighted mean of the profile values across the shell.

It is important to note that this shell spacing will, in general,
lead to shell boundaries that do not line up with any boundaries in the profile
-- whether simply the outer profile edge or
internal boundaries separating layers inside a planet or star.
In the first case of a single-layer profile,
the penultimate particle shell will typically end close to the outer edge.
This leaves a thin and low-mass outermost shell
with only a small number of particles that both
cannot adequately cover the large area
and will be too close in radius to the previous shell.
For interior boundaries such as between core and mantle layers,
a shell will typically straddle the discontinuity.
The particles in this shell then try in vain to represent
some of both materials and conditions.

To avoid these problems, we slightly tweak the input particle mass
to change the mass of the first core shell and hence its radius.
This influences the radii of all the shells (Eqn.~\ref{eqn:dr_shell}).
We iterate the input particle masses
until the boundary of the outermost shell in the first (or only) layer
coincides with the profile's boundary.
This leads to a slightly different total number of particles as well,
but ensures a proper particle representation of the final shell in this layer,
as well as of the first shell of the next layer if there is one.

A similar issue and solution arises for any subsequent boundaries.
To maintain a similar particle mass in all layers,
we do not change the particle mass again.
Instead, we tweak the number of particles
in the first shell of each outer layer.
This changes the mass of that shell and hence its radius, as before.
By using the thickness and density of this shell in Eqn.~\ref{eqn:dr_shell}
instead of the central shell,
this leads to appropriate changes for all the shells in this layer.
We iterate over slightly different numbers of particles in the first shell
until the outermost shell's boundary coincides
with the profile boundary of this layer.
This is repeated at the start of each layer until a particle shell boundary
matches every profile boundary both internally and at the profile's edge.

One remaining decision is at what radius
to place the particles within each shell.
Two average radii to consider are $r_\text{1/2}$,
half-way between the inner and outer radii of the shell,
and $r_\text{m-w}$, the mass-weighted mean radius.
For a slowly changing density profile
and/or many particles that lead to thin shells,
the density is roughly constant throughout the shell and
$r_\text{m-w} > r_\text{1/2}$ because the mass increases with $4\pi r^2$.
In the vast majority of shells, where $dr \ll r$,
these two radii are approximately equal.
However, at small radii near the core,
placing the particles at $r_\text{1/2}$ results in too-high
densities, and $r_\text{m-w}$ gives too-low densities.
We found that placing the particles at
$\tfrac{1}{2} \left(r_\text{1/2} + r_\text{m-w}\right)$
correctly matches the mean SPH density of the particles in each shell
to the profile density at that radius.

\subsubsection{Particles on a Sphere}
\label{sec:method:particles}

For every shell, we now have a number of particles, $N$,
to distribute on the surface of a sphere.
We begin by considering the division of a (unit) sphere
into equal-area regions with small diameters,
following the algorithm described by \citet{Leopardi2007}
with minor modifications.
The particles can then be placed in the centre of each region.

We further impose a stretching of the regions by latitude,
to improve the particle density near the poles.
Finally, each shell is randomly rotated so that the particles at the poles
do not line up in successive shells.

For comparison, we also test the recursive primitive refinement and
parametrised spiral (RPR+PS) method described by \citet{Raskin+Owen2016a}.
Their method uses subdivisions of the Platonic solids for low-$N$ shells
and a spiral placement algorithm for larger numbers of particles.

\begin{figure}
	\centering
	\includegraphics[width=0.9\columnwidth]{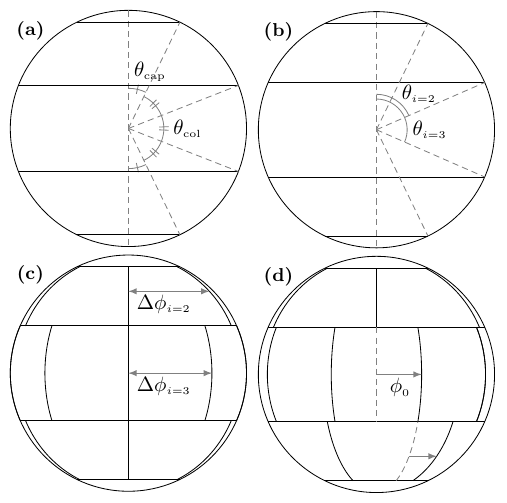}
	\caption{
		An example division of a sphere into 20 equal regions,
		demonstrating the main steps in the algorithm:
		\textbf{(a)} set the polar caps and the initial collar latitudes;
		\textbf{(b)} tweak the collar latitudes so that they each contain
		an integer number of regions;
		\textbf{(c)} divide each collar into equal regions;
		\textbf{(d)} rotate the collars to maximise the minimum separation
		of adjacent regions.
		\label{fig:diagram_EA}}
\end{figure}

\subsubsection{Equal-Area Regions}
\label{sec:eq_area}

\begin{figure}
	\centering
	\includegraphics[width=0.7\columnwidth]{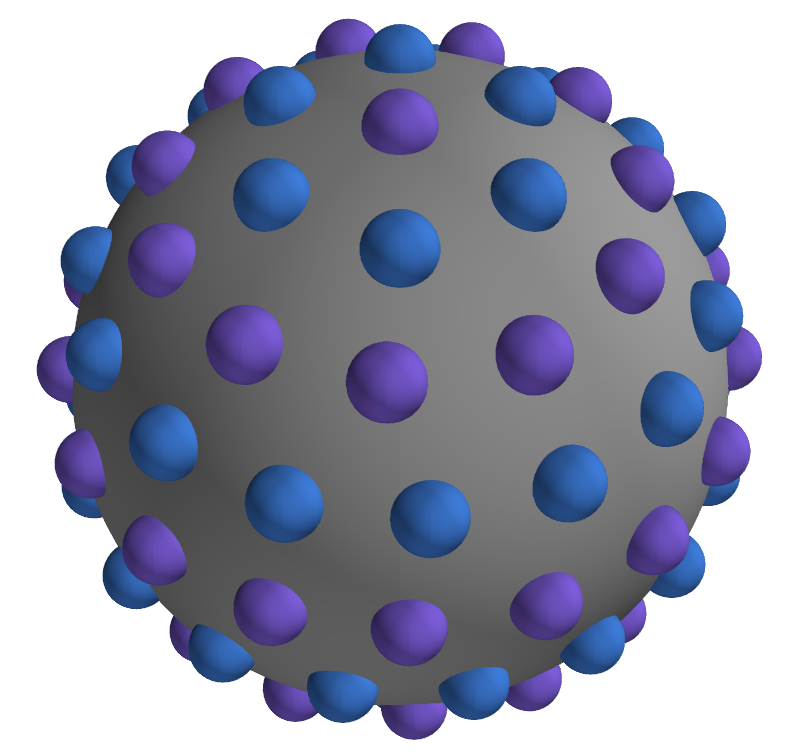}
	\caption{An example of 100 particles distributed on a sphere
		using the SEA (equal-area and subsequent latitude-stretching) method.
		The colours highlight each collar of particles.
		The SPH densities of these particles are shown
		by the purple points in Fig.~\ref{fig:rho_lat}.
		\label{fig:3D_shell_100}}
\end{figure}

This method is also illustrated in Fig.~\ref{fig:diagram_EA}
and a finished example with $N=100$ is shown in Fig.~\ref{fig:3D_shell_100}.

For $N$ regions on a sphere, the area of each one will be
\begin{equation}
	A_{\text{reg}} = 4\pi / N \;.
\end{equation}
The bounding colatitude of a polar cap with area $A_{\text{cap}}$ is
\begin{equation}
	\theta = 2\arcsin\left(\sqrt{\dfrac{A_{\text{cap}}}{4\pi}}\right) \;,
\end{equation}
which for $A_{\text{cap}}=A_{\text{reg}}$ gives the colatitude of
the single-region north pole cap, $\theta_{\text{cap}}$,
and south pole cap, $\pi - \theta_{\text{cap}}$.

We start by dividing the rest of the sphere (between the two polar caps)
into collars with ideal initial heights of $\sqrt{A_{\text{reg}}}$.
This gives the number of collars (when rounded to an integer),
\begin{equation}
	N_{\text{col}} = \text{round}\left[\dfrac{\pi -
		2\theta_{\text{cap}}}{\sqrt{A_\text{reg}}}\right] \;,
\end{equation}
and the actual initial collar height,
\begin{equation}
	\theta_{\text{col}} = \left(\dfrac{\pi -
    2\theta_{\text{cap}}}{N_{\text{col}}}\right) \;,
\end{equation}
(Fig.~\ref{fig:diagram_EA}a).
We then divide each initial collar $i$
into the closest integer number of regions.
The area of each collar is
\begin{equation}
	A_i	= 4\pi \left(\sin^2\left(\dfrac{\theta_i}{2}\right) -
		\sin^2\left(\dfrac{\theta_{i-1}}{2}\right)\right) \;,
\end{equation}
so the ideal number of regions in each collar $i$ is
\begin{equation}
	N'_i = \dfrac{A_i}{A_{\text{reg}}} \;.
\end{equation}
This must be rounded to the actual integer number of regions, $N_i$.
The cumulative discrepancy, $d_i$, from the ideal number of regions
must be included to ensure that the total number of regions is unchanged:
\begin{align}
	N_i 	&= \text{round}\left[N'_i + d_i\right] \\
	d_{i+1} &= d_i + N'_i - N_i \;.
\end{align}
Starting from the north pole and using
the cumulative number of regions in each collar,
$N_{\leq i}$, we find the final colatitude of each collar by
calculating the colatitude of the cap
that contains the same area as $N_{\leq i}$ regions:
\begin{equation}
	\theta_i = 2\arcsin\left(\sqrt{\dfrac{N_{\leq i}
    A_{\text{reg}}}{4\pi}}\right) \;,
\end{equation}
where $i=1$ is the north pole cap (Fig.~\ref{fig:diagram_EA}b).

The points in the centre of each region $j$ in collar $i$ then have
\begin{align}
	\theta	&= \tfrac{1}{2}\left(\theta_i + \theta_{i+1}\right) \\
	\phi	&= \phi_0 + j \, \Delta\phi_i \;,
\end{align}
where $\phi_0$ is the starting longitude
and $\Delta\phi_i = \dfrac{2\pi}{N_{\leq i}}$
is the angle between adjacent points
(Fig.~\ref{fig:diagram_EA}c).

We choose the starting longitude of each collar, $\phi_0$,
to maximise the minimum separation between the points on adjacent collars
(Fig.~\ref{fig:diagram_EA}d).
This helps to prevent local overdensities.
If $N_i$ and $N_{i-1}$ are both odd or both even,
then $\phi_0$ is half the smaller of $\Delta\phi_i$ and $\Delta\phi_{i-1}$.
If one is odd and the other is even,
then $\phi_0$ must be half of the even one's $\Delta\phi$,
to prevent two particles in adjacent collars from having the same $\phi$
and being too close together.

Finally, $\phi_0$ should be additionally offset by $m\,\Delta\phi_{i-1}$,
where $m$ is a random integer between 0 and $N_{i-1}$.
Thus, the $\phi_0$ rotation will be
with respect to a random particle in the previous collar.
This prevents successive collars with large $N_i$
(and hence small $\phi_0$)
from creating a sequence of nearly adjacent particles in successive collars.

\subsubsection{Latitude Stretching}

The equal-area scheme described in \S\ref{sec:eq_area} results in a
small local overdensity of particles near the poles.
We can make the particle density more uniform by stretching the collars near the poles.
However, the collars near the equator must not be overly squashed.
Therefore, the (absolute) latitude of each point, $|\tfrac{\pi}{2} - \theta|$,
should be reduced by an amount that varies with latitude,
from maximum stretching at the poles to 0 at the equator.
Of course, the size of the shift at all latitudes depends on the initial size of the collars,
which is set by the total number of particles.
The collar height and the required shift will decrease in proportion with
the square root of the number of particles.
Thus, the appropriate stretching can be given by:
\begin{equation}
	\theta' = \theta \,+
	\left(\tfrac{\pi}{2} \,-\, \theta\right) \times a N^{-\tfrac{1}{2}} \,
	\exp\left[-\,\dfrac{\tfrac{\pi}{2}\,-\,|\tfrac{\pi}{2}\,-\,\theta|}
	{\pi\,bN^{-\tfrac{1}{2}}}\right] \;,
\end{equation}
where $a = 0.2$ and $b = 2$
(tested for $80 \leq N \leq 10^6$).
For $N<80$, we fit $a$ and $b$ manually
to ensure that the maximum deviation of any particle's density
from the mean is less than $\pm$1\%.
This requires $a$ to vary (non-monotonically) between 0.18 and 0.27,
with $b$ following this variation as $b=10\,a$,
and is only relevant for the innermost one or two lowest mass shells.

\subsection{Planetary Simulations with SWIFT}
\label{sec:methods:swift}

SWIFT (SPH With Inter-dependent Fine-grained Tasking)
is a hydrodynamics and gravity code
for astrophysics and cosmology in open development
(\href{http://swiftsim.com/}{www.swiftsim.com}),
designed from the ground up to run fast and scale well
on shared/distributed-memory architectures \citep{Schaller+2016}.

For the past decade, physical limitations have kept
the speed of individual processor cores constrained,
so instead of getting faster,
supercomputers are getting more parallel.
This makes it ever more important to share the work evenly
between every part of the computer
so that no processors are sitting idle and wasting time.

SWIFT can function as a drop-in replacement for the Gadget-2 code,
which has been widely used for cosmological and planetary impact simulations
\citep{Springel2005,Cuk+Stewart2012},
but with a $>$30$\times$ faster runtime on representative cosmological problems
\citep{Borrow+2018}.
This speed is partly a result of SWIFT's
task-based approach to parallelism and domain decomposition
for the gravity and SPH calculations \citep{Gonnet2015}.
By evaluating and dividing up the work instead of just the data,
this provides a dynamic way
to achieve good load balancing across multiple processors
within a shared-memory node.
The tasks are decomposed over the network in distributed memory systems
using a graph-partitioning algorithm,
weighting each task by the estimated computational work it requires.
Combined with using asynchronous communications
that are themselves treated as normal tasks,
this allows the code to scale well \citep{Schaller+2016}.
Core routines, including the direct interaction between particles,
have then been optimized using vector instructions \citep{Willis+2018}.

In some respects, giant impact simulations
pose a harder challenge for load balancing
than the cosmological simulations that SWIFT is also designed for.
For a large patch of the universe,
although the density becomes very much higher in a galaxy than a void,
the local average density is roughly constant across a simulation box.
Even a crude division of particles by position in the box
to different computing cores can somewhat effectively speed up the calculation,
and a more careful decomposition like SWIFT's
can produce excellent strong scaling
across hundreds of thousands of cores \citep{Borrow+2018}.

In contrast, for a giant impact,
almost all the mass (and hence particles) is in the planet at the centre.
If we use a large simulation box in order to follow the ejected debris,
then the vast majority of particles can easily occupy
less than 0.01\% of the volume.
This is similar to cosmological `zoom-in' simulations
that use a high-resolution region to focus on a single galaxy or halo.
This firstly makes it harder to divide up particles between computing nodes,
and secondly can require much more frequent communication.
This makes it much less efficient to use a large numbers of cores,
and difficult to fully utilise a large supercomputer to run
a single planetary simulation very quickly.

Happily, most studies of giant impacts can be reframed as
`embarrassingly parallel' problems because,
instead of investigating one specific collision in extreme detail,
the usual aim is to study a wide range of scenarios,
such as varying the impact angle and speed.
For this reason, perfect scaling across many distributed-memory nodes
or MPI ranks is not as important.
Many impacts can each be simulated on their own single
(or small number of) shared-memory node(s).
SWIFT then uses threads and SIMD vectorisation
to parallelise efficiently across the tens of cores within each node.
However, as we investigate in \S\ref{sec:results:uranus}, 
even for parameter-space surveys,
large numbers of particles may be necessary 
to obtain sufficiently converged results,
depending on the property being studied.

\subsubsection{Planetary SPH}
\label{sec:methods:planetary_sph}

SWIFT has a modular structure that separates different code sections
for clean modifications to, for example, the physics or the hydrodynamics scheme
without affecting (or even being aware of)
the parallelisation and other structural components.
Any such module is switched in or out with configuration flags,
allowing SWIFT to run planetary, cosmological,
or any other simulation as required.

The hydrodynamics scheme used for the simulations in this paper
uses a simple `vanilla' form of SPH as described in e.g. \citet{Price2012},
with the Balsara switch for the artificial viscosity \citep{Balsara1995}.
Multiple other schemes are also implemented in SWIFT,
as well as various SPH kernels.
Here, we use the simple 3D cubic spline kernel with 48 neighbours,
corresponding to a ratio of smoothing length to inter-particle separation
of $\gamma = 1.2348$ \citep{Dehnen+Aly2012}.
The default artificial viscosity parameters
for the \citet{Monaghan1992} model
are set to $\alpha=1.5$ and $\beta = 2\,\alpha$,
as is typical in the literature \citep[e.g.][]{Reinhardt+Stadel2017}.

The equation of state (EoS) for a material relates its pressure
to its density and temperature (or internal energy or entropy).
So far,\footnote{
  The simulations in this paper used SWIFT version 0.8.1.
  }
we have implemented several
Tillotson, SESAME, and \citet{Hubbard+MacFarlane1980}
(for Uranus materials) EoS in SWIFT,
as well as an ideal or isothermal gas.
Any number of these different materials can be simulated together,
as is required in a multi-layered planet, for example.

\section{Results and Discussion}
\label{sec:results}

\subsection{Particle Placement}
\label{sec:results:particles}

In this section, we first test the arrangement of particles
on an isolated spherical shell.
Then, we investigate full 3D initial conditions for a simple Earth-mass planet,
considering the SPH densities of the particles in their initial positions
and how close they are to equilibrium when allowed to evolve.

\begin{figure}
	\centering
	\includegraphics[width=\columnwidth, trim={6mm 8mm 9mm 9mm}, clip]{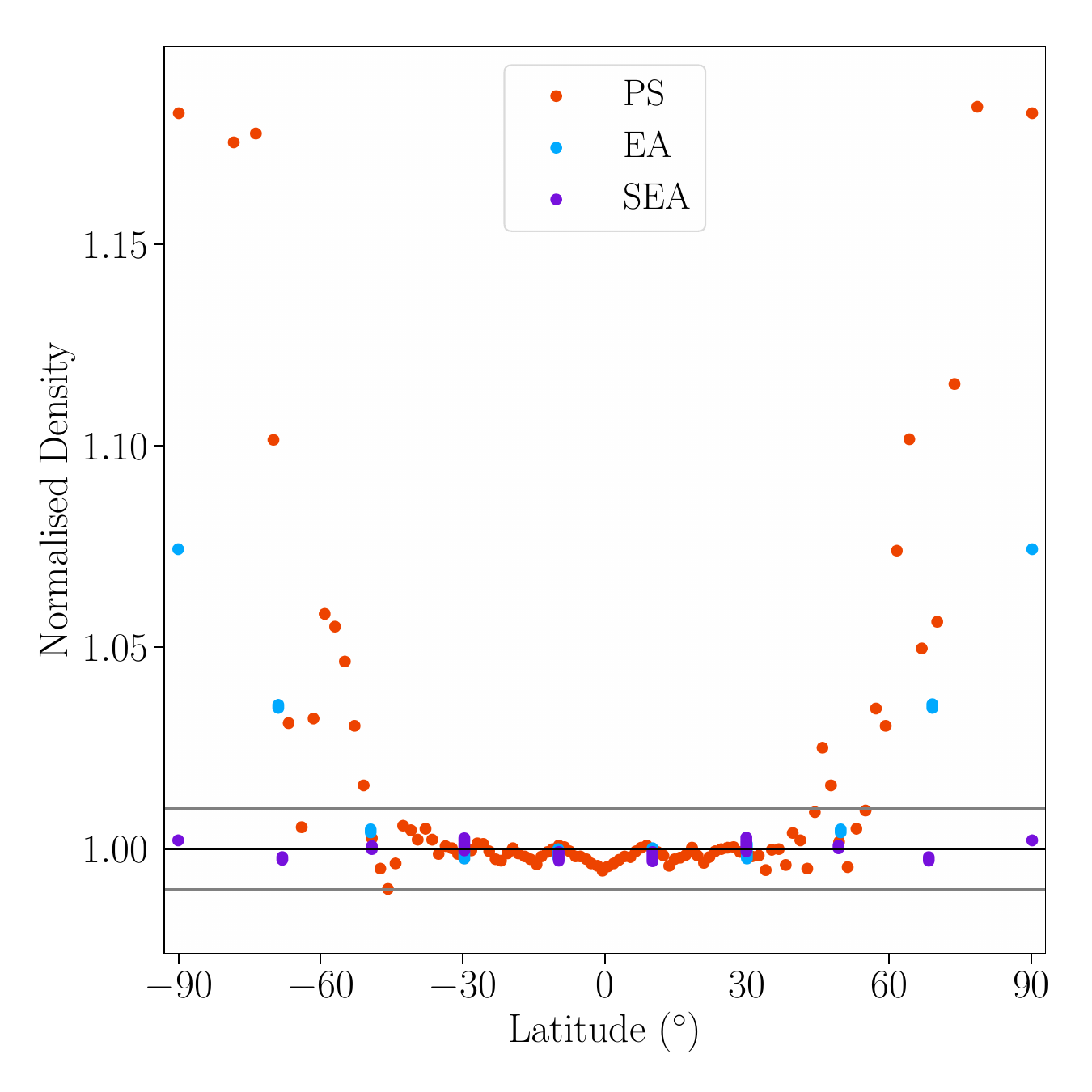}
  \vspace{-1.5em}
	\caption{The SPH densities of 100 particles placed using
		three different schemes,
		normalised by the median density.
		The grey lines show $\pm1$\% of the median.
		The 3D positions of these SEA particles are illustrated in Fig.~\ref{fig:3D_shell_100}.
		\label{fig:rho_lat}}
\end{figure}

Fig.~\ref{fig:rho_lat} shows the densities of 100 particles arranged
on a unit spherical shell using three different methods:
\citet{Raskin+Owen2016a}'s recursive primitive refinement
and parametrised spiral method (RPR+PS, specifically PS in this case)
and our equal-area method without (EA)
and with (SEA) the extra latitude stretching,
as described in \S\ref{sec:method:particles}.

\begin{figure}
	\centering
	\includegraphics[width=\columnwidth, trim={6mm 8mm 9mm 9mm}, clip]{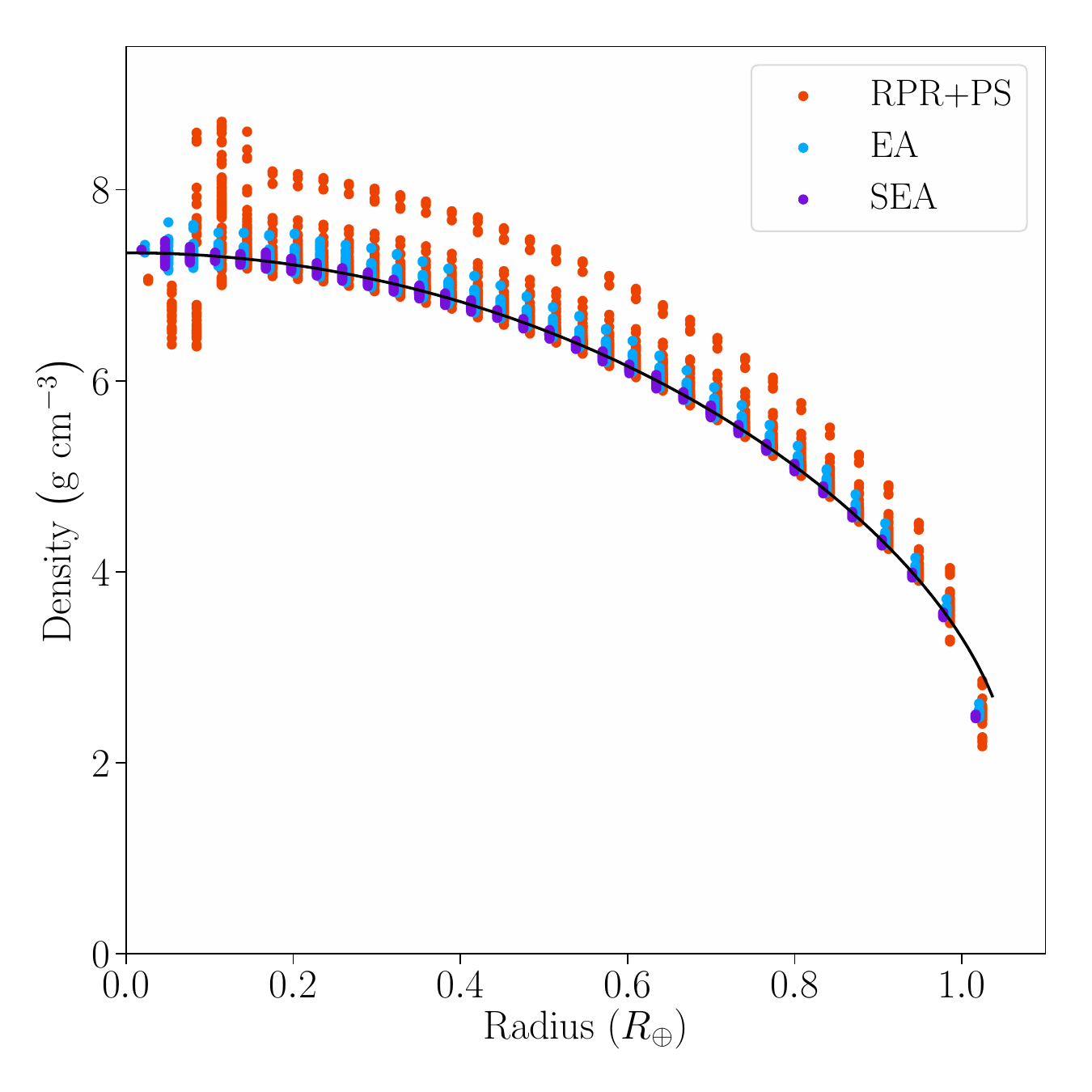}
  \vspace{-1.5em}
	\caption{The SPH densities of $\sim$$10^5$ particles placed using
		the three different shell schemes as labelled in the legend.
    The EA and RPR+PS particles are shown
    offset to slightly higher radii for clarity.
		The black line shows the input density profile,
		representing a simple model of an Earth-mass planet.
    The SEA particles' densities stay within 1\% of the profile,
    as in the Fig.~\ref{fig:rho_lat} isolated shell case.
		\label{fig:rho_prof_init}}
\end{figure}

\begin{figure*}
	\begin{center}
		\includegraphics[width=\textwidth, trim={12mm 110mm 13mm 13mm}, 
    clip]{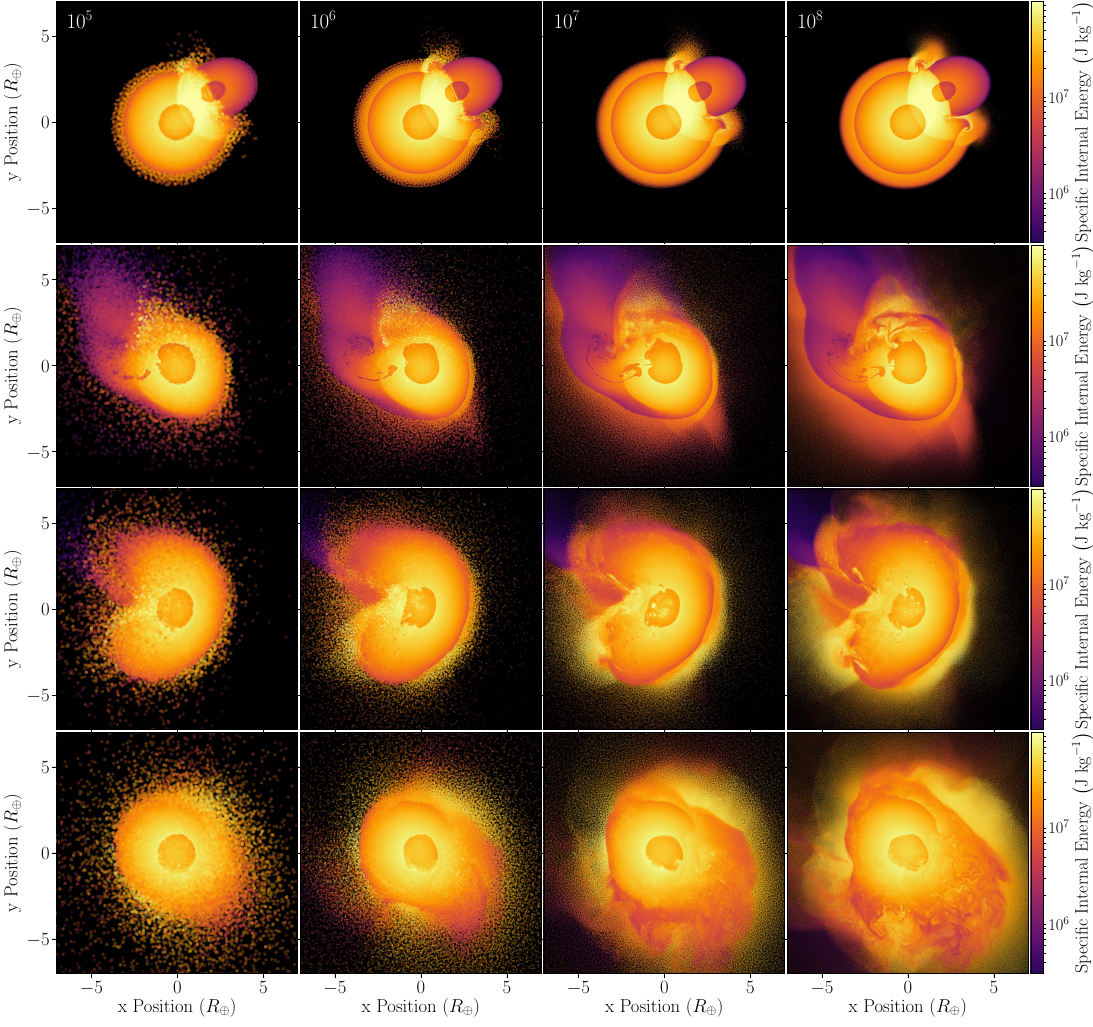}
		\vspace{-2.5em}
		\caption{
      Mid-collision snapshots in the early stages of
      the same giant impact on Uranus
      at the same times from simulations with
      the $\sim$$10^5$ SPH particles (left panels) typical in the literature,
      up through $10^6$ and $10^7$ to the $10^8$ (right panels)
      made possible with SWIFT,
      resolving more of the detailed evolution of
      both internal structure and debris.
      Snapshots shown are $\sim$2, 3, 4, and 7 hours 
      after the start of the simulation.
      An animation of the highest resolution impact is available at
      \href{http://icc.dur.ac.uk/giant_impacts/uranus_1e8_anim.mp4}{icc.dur.ac.uk/giant\_impacts}.
      }
		\label{fig:uranus_snaps}
	\end{center}
	\vspace{-2em}
\end{figure*}

The RPR+PS and EA methods both show significant overdensities at the poles,
with maximum deviations from the median density
approaching 20\% and 10\% respectively.
This is still a big improvement on a random distribution of particles
on a shell, which leads to densities that are wrong by a factor of $>$10.
The SEA stretching reduces the scatter to less than 1\%,
with typical maximum deviations of 0.5\%,
depending on the exact number of particles.
Only 100 particles are shown here for clarity;
the three methods show similar relative deviations
for $10^2$--$10^6$ particles in a single shell.

Unfortunately, this dramatic improvement of SEA over the unstretched EA method
cannot be replicated for RPR+PS because the distribution of particles
is not azimuthally symmetric.
Stretching the RPR+PS particles at the poles
reduces the overdensity for some particles
but creates unavoidable underdensities for others because of their asymmetry.

To investigate how these properties of an isolated shell
translate into nested shells in 3D,
we now consider a full model of an Earth-mass planet with $\sim$$10^5$ particles
(see appendix~\ref{sec:profiles}).
The results from using the same three placement methods
are shown in Fig.~\ref{fig:rho_prof_init}.
As in the isolated-shell case,
the RPR+PS particles show a large range of densities,
with a systematic spread of particle densities
more than 10\% discrepant from the profile.
The unstretched EA method shows similar density discrepancies around 4\%,
while the SEA stretching again ensures the scatter
is within 1\% of the profile density.
These values are for a cubic spline kernel with 48 neighbours.
Using another common example of the Wendland-C6 kernel with 200 neighbours
yields the same qualitative results but
reduces the density scatter in all cases by roughly $\tfrac{1}{2}$.

The underdensity of particles in the outermost shell
is caused by the nature of the SPH density calculation,
so is seen equally for all methods.
The spherical kernel volume extends into
the empty space above the planet's surface without finding any neighbours,
artificially reducing the density.

It is noteworthy that the density deviations of
the RPR+PS and EA methods were reduced
when switching from the 2D to the 3D case,
while the SEA deviations were approximately unchanged.
This reflects the contributions of
the particles in other shells to the SPH density.
The high overdensities are reduced in 3D because
the nearby particles in adjacent shells are also summed over,
mitigating the impact of the too-close particles in the same shell.
For SEA, the particles in the randomly rotated adjacent shells
are just as likely to be very slightly too close or too far
as the particles in the same shell,
so the density discrepancies are largely unchanged.
This suggests that there would be little benefit to improving the
distribution of particles within each shell beyond that of SEA,
e.g. by running a relaxing simulation within each shell.
Even if the particles in every isolated shell were perfectly arranged,
then the imperfect contributions from adjacent-shell particles
would negate any improvement.
So, if even smaller density deviations were desired,
then it would be necessary to consider all particles at once.

The actual success of our method is determined by
how close the particles are to equilibrium
when allowed to evolve in a simulation.
A standard criterion for initial conditions to be considered `relaxed'
enough for use is that the root mean square velocity, $v_{\rm rms}$,
is below $\sim$1\% of the escape speed, here $v_{\rm esc} = 11.2$~km~s$^{-1}$.
Thanks to their precise densities,
the SEA particles immediately have $v_{\rm rms}$ below 0.01~$v_{\rm esc}$,
and the maximum particle speed first peaks at under 0.04~$v_{\rm esc}$.
(`Immediately' here meaning the fastest speeds the particles reach,
soon after being allowed to evolve from a stationary start.)
In comparison, a random distribution of particles in shells
has initial $v_{\rm rms} = 0.2~v_{\rm esc}$.

\begin{figure*}
	\begin{center}
		\includegraphics[width=\textwidth, trim={13mm 194mm 14mm 17mm}, clip]{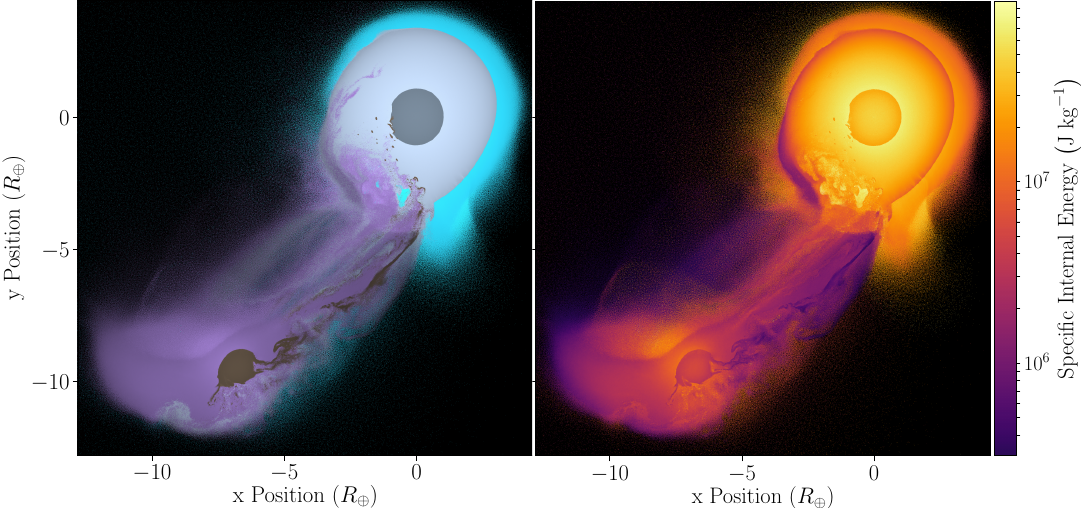}
		\vspace{-2em}
		\caption{
      A mid-collision snapshot of a grazing impact with $10^8$ SPH particles
      -- compared with the more head-on collision in
      Fig.~\ref{fig:uranus_snaps} --
      coloured by their material and internal energy,
      showing some of the detailed evolution and mixing
      that can now be resolved.
  		In the left panel,
      light and dark grey show the target's ice and rock material, respectively,
  		and purple and brown show the same for the impactor.
  		Light blue is the target's H-He atmosphere.
      }
		\label{fig:uranus_grazing}
	\end{center}
	\vspace{-2em}
\end{figure*}

Most of the SEA particles' motion is caused by the previously mentioned
underdensity of the outermost shell,
which causes the entire planet to gently oscillate
and settle into a slightly lower density profile.
Because this dominates the discrepancy from an equilibrium state,
the RPR+PS particles' $v_{\rm rms}$ is almost identical to SEA
in spite of their comparatively noisy densities.
Their maximum speed is slightly higher at 0.07~$v_{\rm esc}$.
If a modified density estimator is used to fix the outer boundary problem,
then a larger difference might be expected between the two methods.
Planets with layers of different materials --
such as the proto-Uranus and impactor in \S\ref{sec:results:uranus} --
face similar SPH density problems at interior boundaries as well.

We confirmed that these relaxed SEA results are unchanged
for Moon- and Pluto-mass planets ($\sim$0.01 and 0.002~$M_\oplus$),
which are less strongly gravitationally bound,
making them slightly less stable.
However, the Tillotson EoS used here \citep{Tillotson1962}
is even steeper close to the low density at which the pressure is zero,
as is the case for other EoS and depending on the temperature.
This exacerbates any density errors into even greater pressure discrepancies.
For RPR+PS, some under-dense particles in the Pluto-mass planet
are even pushed below the zero-pressure density,
while the most over-dense ones get assigned a pressure
over 4 times the desired value.
Nevertheless, these particles can quickly be relaxed
without much affecting the overall structure or $v_{\rm rms}$.
SEA has the mild advantage that it avoids such issues in the first place,
and requires similarly trivial computation to generate the initial conditions.

The SEAGen code for quickly generating both isolated shells
and full spheres of points is publicly available at
\href{https://github.com/jkeger/seagen}{github.com/jkeger/seagen}
or can be installed directly with pip as the python module \verb|seagen|.

\subsection{Uranus Giant Impacts and Convergence}
\label{sec:results:uranus}

We now use these tools for first creating and then simulating planets
to study the convergence (or lack thereof) 
of giant impact simulations using $10^5$ up to $10^8$ SPH particles.
We focus on three science-motivated questions 
about the giant impact that likely knocked over 
the planet Uranus to spin on its side:
(1) How much atmosphere is ejected from the system? 
(2) How much rocky material is placed into orbit? 
(3) What is the post-impact rotation period of the planet? 

Here, we repeat some of the simulations from \citetalias{Kegerreis+2018} \citep{Kegerreis+2018}
with $\sim$$10^5$, $10^6$, $10^7$, and $10^8$ particles
to investigate how these higher resolutions compare 
with the current standard,
and to demonstrate the simulation tools described in this paper.
The full details of the equations of state and initial conditions
are described in \citetalias{Kegerreis+2018}.

Fig.~\ref{fig:uranus_snaps} shows comparisons of
a typical impact simulated at different resolutions,
repeating the `low angular momentum' scenario of \citetalias{Kegerreis+2018}'s Fig.~2.
Although the overall behaviour is encouragingly similar,
details like the tidal stretching of the impactor's core
and the distribution of the debris
clearly cannot be fully resolved by the $10^5$ or $10^6$ particle simulations.
Fig.~\ref{fig:uranus_grazing} highlights some the details
that can be resolved with $10^8$ particles
for the grazing impact of the
`high angular momentum' scenario of \citetalias{Kegerreis+2018}'s Fig.~3.

\begin{figure*}
	\begin{center}
		\includegraphics[width=\textwidth, trim={52mm 9mm 74mm 9mm}, clip]{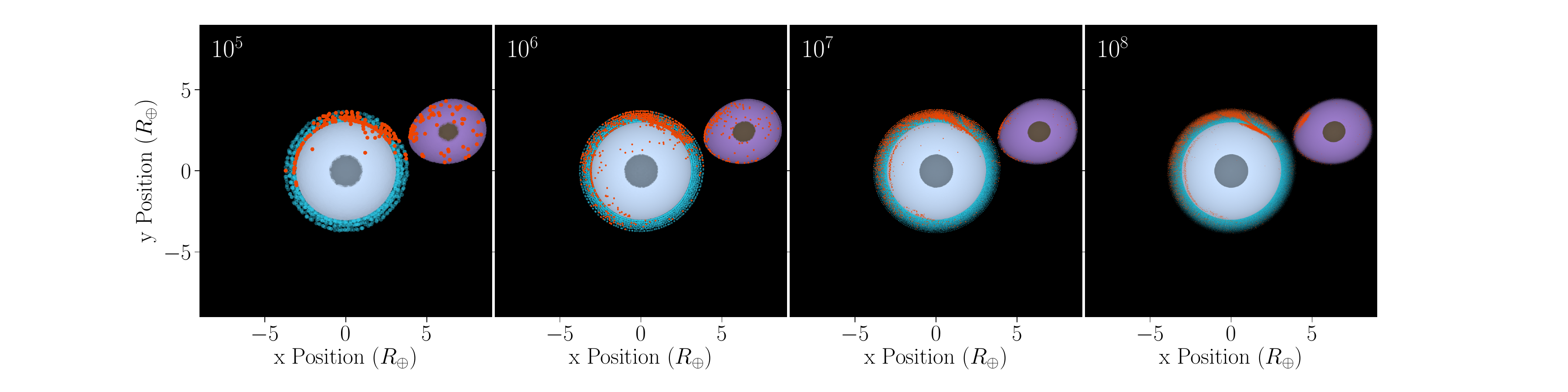}
		\vspace{-1.5em}
		\caption{
      The particles that will become unbound and escape the system,
      highlighted in orange on a pre-impact snapshot
      from the same simulations with $\sim$$10^5$--$10^8$ SPH particles
      as in Fig.~\ref{fig:uranus_snaps}.
      Only particles in a thin cross-section are shown for clarity.
      The colours are the same as in Fig.~\ref{fig:uranus_grazing}.
      The times at which these particles become unbound 
      are shown in Fig.~\ref{fig:unbound_evol}.
      The total mass lost is broadly similar in all cases,
      but $10^5$--$10^6$ particles fail to resolve the detailed results.}
		\label{fig:uranus_unbound}
	\end{center}
	\vspace{-2em}
\end{figure*}

\subsubsection{Ejected Debris}
\label{sec:uranus:debris}

In \citetalias{Kegerreis+2018} we found that the majority of the atmosphere survives the impact,
but that a small fraction can be fully ejected.
Fig.~\ref{fig:uranus_unbound} highlights the particles
that will become gravitationally unbound and escape from the system.
The initial collision blasts away much of the outer atmosphere and some ice,
some of which will escape but most remains gravitationally bound.
The $10^7$ and $10^8$ particle runs show that
a deeper shell of now-exposed particles
then gets ejected during the subsequent violent oscillations
as the impactor remnants fall back in and the planet slowly starts to settle.

\begin{figure}
	\begin{center}
		\includegraphics[width=\columnwidth, trim={5mm 5mm 5mm 5mm}, clip]{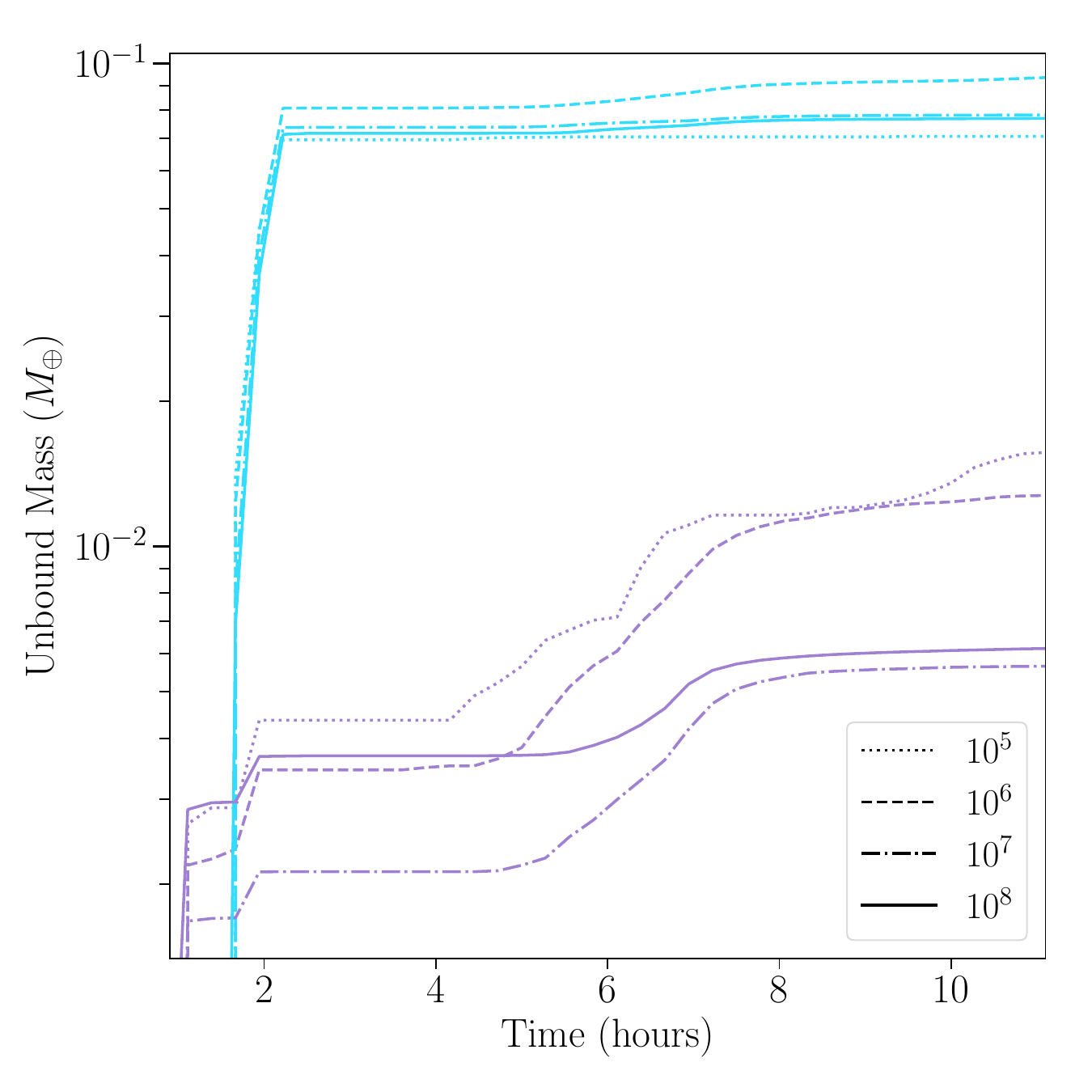}
		\vspace{-1.5em}
		\caption{
      The time evolution of the mass of 
      gravitationally unbound atmosphere (light blue) 
      and impactor-ice (purple) material that is ejected from the system
      -- the same particles highlighted in Fig.~\ref{fig:uranus_unbound} --
      for the different resolution simulations.
      }
		\label{fig:unbound_evol}
	\end{center}
  \vspace{-0.5em}
\end{figure}

The time at which this ejected material becomes unbound 
in each simulation is shown in Fig.~\ref{fig:unbound_evol}.
Significant mass is blasted off the planet 
even several hours after the initial collision in all cases.
The $10^7$ and $10^8$ simulations closely agree
that 9\% of the total atmosphere mass escapes.
The $10^5$ and $10^6$ simulations differ 
(non-systematically) with 8\% and 12\%, respectively.
This suggests that atmospheric erosion has converged 
by $\sim$$10^7$ particles in this case.
On the positive side, although the lower resolution simulations 
do not show perfectly converged behaviour, 
for answering the practical question of how much atmosphere is lost,
all simulations give a qualitatively similar answer of $\sim$10\%.

Most studies of impact erosion use analytical or one-dimensional models
to estimate the ejected atmosphere given a certain ground speed 
from the shock induced by the impact \citep[e.g.][]{Inamdar+Schlichting2016}.
In this case, the initial shock removes 8\% of the atmosphere,
then an additional 1\% is lost in the subsequent sloshing.
So, much like the minor resolution dependence, 
general conclusions about the fraction of atmosphere lost to an impact 
of this scale are unlikely to change.
However, for more precise studies, smaller atmospheres, 
and perhaps other impact scenarios,
this process should not be ignored.

For comparison, also shown in Fig.~\ref{fig:unbound_evol} 
is the mass of unbound ice. 
The $10^7$ and $10^8$ simulations again give similar final answers,
but do not show the same behaviour at earlier times.
The lower resolution simulations are discrepant by more than a factor of 2.
It seems plausible that this quantity is approaching convergence,
but without more particles than $10^8$ (or checking $10^{7.5}$),
it is clearly not safe to assume this is a fully reliable result.

\begin{figure}
	\begin{center}
		\includegraphics[width=\columnwidth, trim={5mm 8mm 5mm 5mm}, clip]{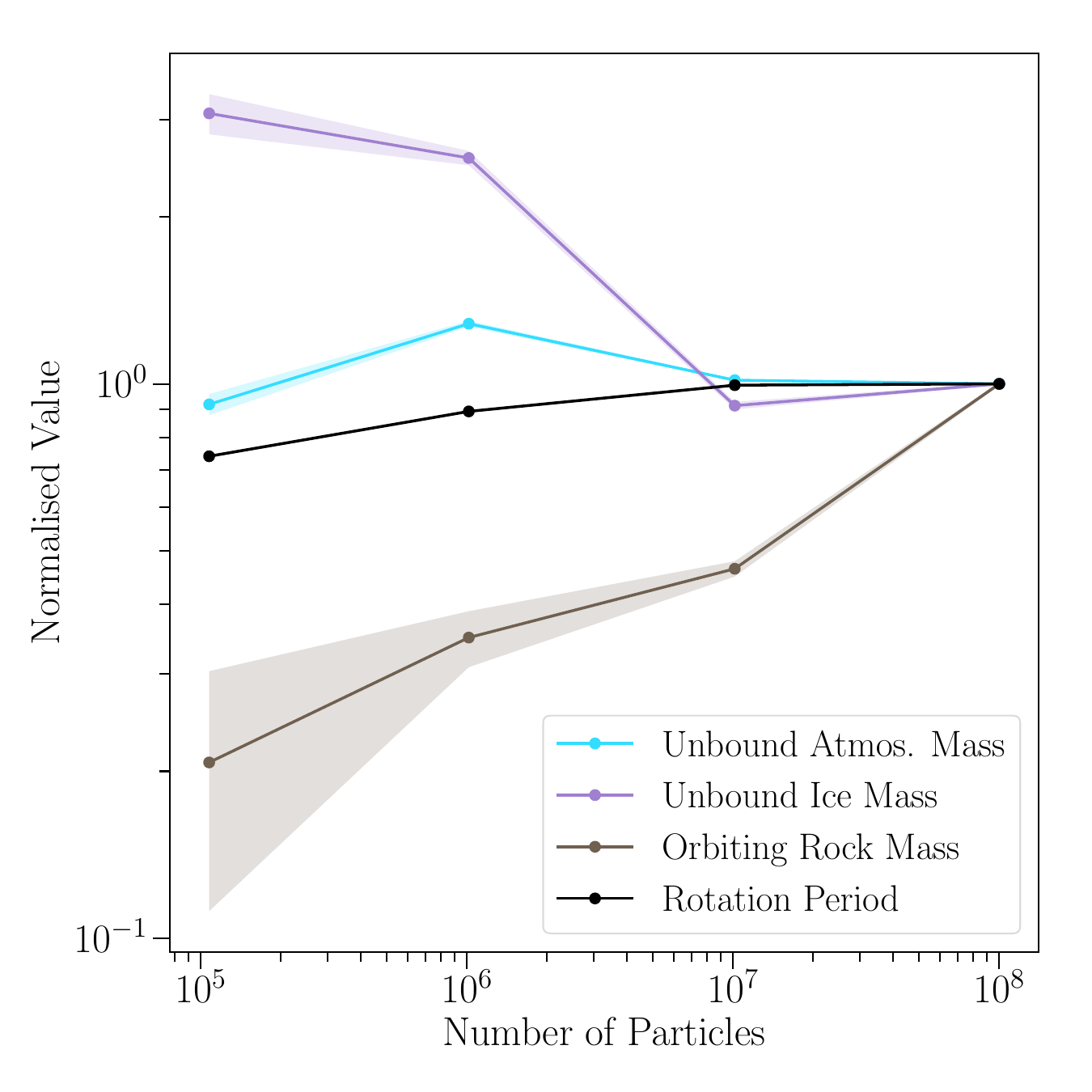}
		\vspace{-1.5em}
		\caption{
      The change with resolution of 
      the masses of unbound atmosphere and ice,
      the mass of rock placed into orbit beyond the Roche radius,
      and the planet's rotation rate,
      demonstrating a range of apparent (un)convergence.
      Each property is normalised by the highest-resolution result
      to show the relative differences.
      The shaded regions show the 1-$\sigma$ errors,
      some of which are too narrow to see.
      The rotation period appears to have converged by $\sim$$10^7$ particles,
      as have -- with decreasing certainty -- 
      the unbound atmosphere and ice masses,
      while the orbiting rock mass has not.
      }
		\label{fig:compare_resolution}
	\end{center}
  \vspace{-0.5em}
\end{figure}

These quantities are summarised in Fig.~\ref{fig:compare_resolution}
at 14~hours as a function of the number of particles,
showing by how much each simulation differs from the highest resolution.
That the eroded atmosphere appears closer to convergence than the ice 
is not surprising given the order-of-magnitude lower mass of ejected ice,
meaning correspondingly fewer particles 
are involved in attempting to resolve the process -- 
as can be interpreted by the size of the error bars.

As an example of a property that has certainly not converged, 
we also plot the mass of rock that is ejected into orbit in a debris disk
beyond the Roche radius, 
where it might be available for accretion into satellites.
Not only do the $10^7$ and $10^8$ simulations not agree,
they differ by more than the lower resolutions
with no semblance of convergence.
The corresponding number of orbiting rock particles in each simulation 
is only 4, 80, 1000, and 20,000, respectively.
So, especially for a messy ejection process 
that is widely spread out in both space and time,
it is not surprising that 1000 or fewer particles 
are far from able to sufficiently resolve what happens.
It is possible that the $10^8$ simulation 
has already fully resolved and converged on this result,
but our only means of checking this
-- running even higher resolution simulations --
we leave for future studies where this is a targeted science result.
In comparison, the masses of orbiting ice and atmosphere particles
in the debris are much higher,
and converge similarly to the unbound atmosphere mass.

\subsubsection{Rotation Period}
\label{sec:uranus:rotation}

The rotation period of the post-impact planet 
is a large-scale bulk property involving a large majority of all particles, 
so one might expect it to have converged by fairly low particle numbers.
However, as shown in Fig.~\ref{fig:compare_resolution},
while the $10^7$ and $10^8$ simulations agree on a rotation period of 19.9~hours
to within 0.5\% of each other,
the $10^5$ and $10^6$ simulations find 
much shorter periods of 14.7 and 17.7~hours.
This is a significant change to our results in \citetalias{Kegerreis+2018},
when it appeared that even fairly low-impact-parameter 2~$M_\oplus$ 
impactors could impart enough spin to explain the planet today.
Assuming an approximately similar reduction in spin for other impact scenarios,
only a narrower range of more-grazing impacts
(or more massive impactors) would be viable.

\begin{figure}
	\begin{center}
		\includegraphics[width=\columnwidth, trim={5mm 5mm 5mm 5mm}, clip]{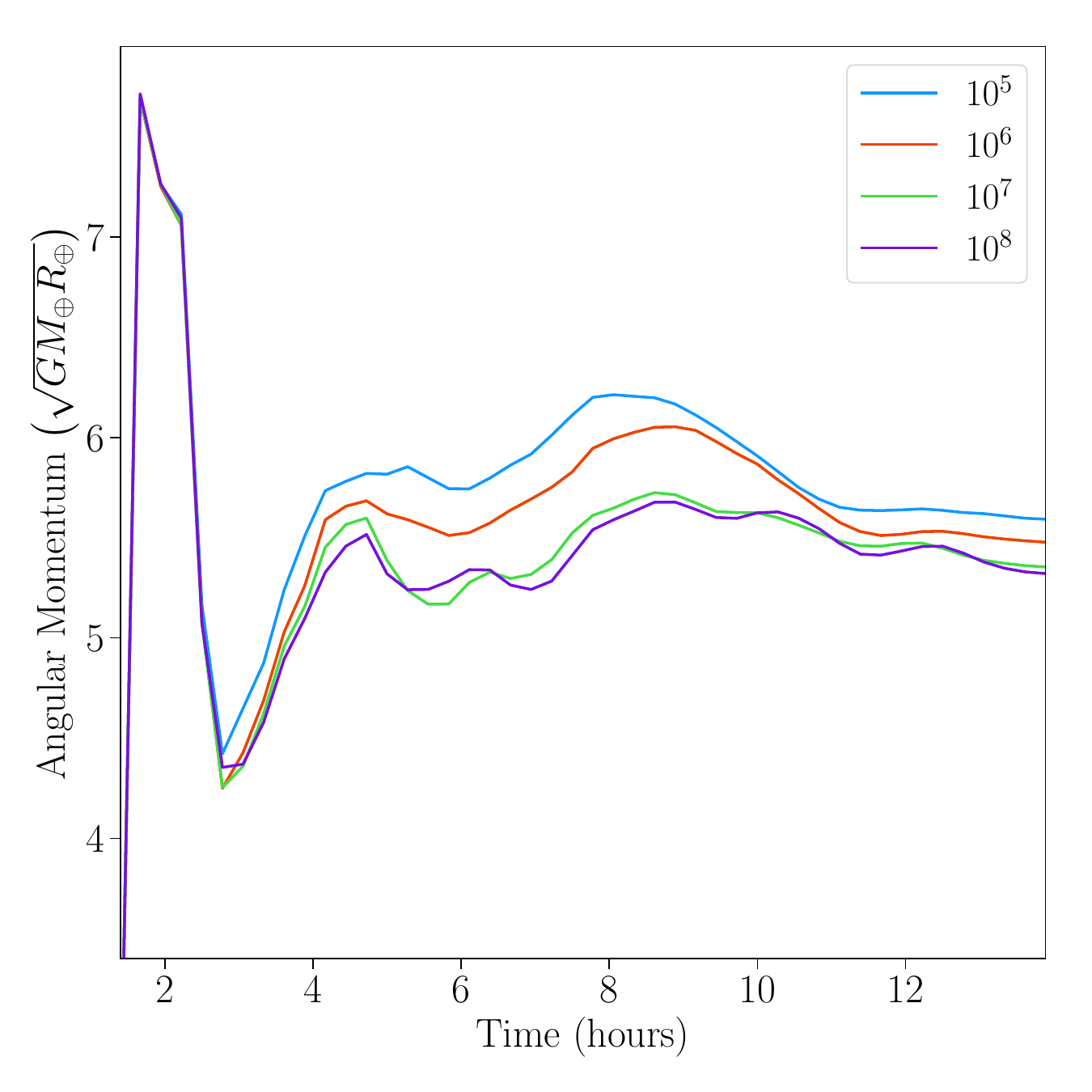}
		\vspace{-1.5em}
		\caption{
      The early time evolution of the planet's angular momentum
      for the different resolution simulations,
      summed over all particles within the Roche radius of 6~$R_\oplus$.
      The standard- and high-resolution simulations begin to differ 
      as the ejecta from the initial impact falls back to the planet.
      }
		\label{fig:ang_mom_evol}
	\end{center}
  \vspace{-0.5em}
\end{figure}

The evolution of the planet's angular momentum for each simulation 
is shown in Fig.~\ref{fig:ang_mom_evol},
which, for simplicity, we sum over all particles within the Roche radius.
The total angular momentum of the entire system remains the same in all cases,
but at higher resolutions more angular momentum is transported out 
to the debris disk beyond the Roche radius,
leaving less in the planet.
All the simulations agree during the arrival 
and initial merging of the impactor,
but their behaviour begins to diverge as the thrown-out debris 
(see the middle two rows of Fig.~\ref{fig:uranus_snaps})
begins to fall back in to the planet, 
at around 3~hours after the start of the simulation.

Even though the total number of particles used to measure 
the planet's rotation rate is very large,
the messy ejecta and mixing around the outer regions of the planet
are significant enough to affect the overall system
while also small enough to require high resolutions to model correctly.
This is comparable to the effect seen by \citet{Hosono+2017}
where the mass of the post-impact disk 
did not converge as expected
because of subtle differences in the detailed behaviour of re-impacting debris.

There will always be even smaller structures that are not properly resolved,
but their ability to alter the rest of the system will also decrease,
so appropriate-scale quantities should stay converged.
However, properties such as small-scale turbulent mixing and 
the emergence of smaller structures may never converge 
without the addition of regularising physics 
such as diffusion or viscosity mechanisms
\citep{Lecoanet+2016,Cullen+Dehnen2010}.

On the convergence of the rotation rate, 
in addition to the similar angular momenta 
of $10^7$ and $10^8$ throughout time,
the rotation period encouragingly changes monotonically with higher resolution 
and by less with each increase.
So we interpret the various quantities 
shown in Fig.~\ref{fig:compare_resolution}
as demonstrating a range of behaviour 
from the apparently well-converged rotation rate and unbound atmosphere mass
by $10^7$ particles,
through the possibly converged unbound ice,
to the clearly un-converged orbiting rock.

\section{Conclusions}
\label{sec:conclusions}

We have presented a simple method for creating
spherical arrangements of particles with precise densities,
and the SWIFT code for hydrodynamical simulations,
then used them to study giant impacts at high resolutions.

The SEA algorithm allows the quick creation of
near-equilibrium, spherically symmetric initial conditions of particles
(\href{https://github.com/jkeger/seagen}{github.com/jkeger/seagen}).
It ensures that every particle has an SPH density
within 1\% of the desired value,
unlike the otherwise-similarly successful methods of
\citet{Raskin+Owen2016a} and \citet{Reinhardt+Stadel2017}.
This mitigates the need for expensive computation
that is otherwise required to produce initial conditions
that are relaxed and ready for a simulation.

The open-source SWIFT code is designed 
to take advantage of contemporary shared/distributed-memory architectures
(\href{http://swiftsim.com/}{www.swiftsim.com}).
For planetary giant impact simulations,
this has enabled a 100--1000$\times$ improvement
in the number of particles that can be used,
allowing the study of brand new topics that were out of reach for
lower resolution simulations.

To demonstrate these tools and test the convergence of such simulations,
we revisited the study of the giant impact onto the young Uranus
that may explain its spin and other strange features \citep{Kegerreis+2018}.
We find that even large-scale results such as the rotation rate 
are not converged with standard-resolution simulations 
of $10^5$ and $10^6$ particles.
The overall behaviour is similar in all cases, 
but small variations in the debris that falls back after the initial impact 
have a significant effect on the post-impact planet and its rotation rate,
which appears to be well-converged with $10^7$ and $10^8$ particles,
but not fewer.
Similar but mildly less certain convergence is seen for the 
masses of atmosphere and ice that are ejected from the system,
while the low mass of rock placed into orbit has not converged at all
by $10^7$ particles.

Increasing resolution is only one important challenge 
for developing more realistic simulations.
We have here used a simple implementation of SPH
with a focus on simply increasing the number of particles.
Future studies must continue to test high resolutions with, for example, 
more sophisticated equations of state
and improved SPH formulations with better treatment of
issues such as material and density discontinuities.

We conclude that standard-resolution simulations with $<$$10^7$ SPH particles 
can fail to produce reliable results 
even for large-scale properties of planetary system.
$10^7$ and $10^8$ particles appear to pass the threshold 
of resolving the major processes in a giant impact.
However, different collisions and other specific simulation outputs
will depend more or less strongly on the behaviour of smaller structures,
with correspondingly different requirements for convergence.
The highly non-linear nature of giant impacts
and the combinations of short- and long-term, 
localised and distributed processes
prevent simple predictions for how many particles 
will be sufficient for a given result to converge.

\section*{Acknowledgements}

We thank James Willis, Josh Borrow, and all the members of the SWIFT team,
and thank Lydia Heck for invaluable computational advice and support.
We thank the anonymous referee for their constructive comments.
The research in this paper made use of the SWIFT open-source simulation code
\citep[\href{http://www.swiftsim.com}{www.swiftsim.com},][]{Schaller+2018} 
version 0.8.1.
This work used the DiRAC@Durham facility managed by the Institute for
Computational Cosmology (ICC) on behalf of the STFC DiRAC HPC Facility
(\href{www.dirac.ac.uk}{www.dirac.ac.uk}).
The equipment was funded by BEIS capital funding
via STFC capital grants ST/K00042X/1, ST/P002293/1, ST/R002371/1 and
ST/S002502/1, Durham University and STFC operations grant
ST/R000832/1. DiRAC is part of the National e-Infrastructure.
This work was supported by INTEL through establishment of the ICC
as an INTEL parallel computing centre (IPCC).
JAK is supported by STFC grant ST/N50404X/1
and the ICC PhD Scholarships Fund.
VRE acknowledges support from STFC grant ST/P000541/1.
RJM is supported by the Royal Society.
MS is supported by VENI grant 639.041.749.
LFAT acknowledges support from NASA Outer Planets Research Program
award NNX13AK99G.




\bibliographystyle{mnras}
\bibliography{../gihr.bib}

\begin{thebibliography}{}
\makeatletter
\relax
\def\mn@urlcharsother{\let\do\@makeother \do\$\do\&\do\#\do\^\do\_\do\%\do\~}
\def\mn@doi{\begingroup\mn@urlcharsother \@ifnextchar [ {\mn@doi@}
  {\mn@doi@[]}}
\def\mn@doi@[#1]#2{\def\@tempa{#1}\ifx\@tempa\@empty \href
  {http://dx.doi.org/#2} {doi:#2}\else \href {http://dx.doi.org/#2} {#1}\fi
  \endgroup}
\def\mn@eprint#1#2{\mn@eprint@#1:#2::\@nil}
\def\mn@eprint@arXiv#1{\href {http://arxiv.org/abs/#1} {{\tt arXiv:#1}}}
\def\mn@eprint@dblp#1{\href {http://dblp.uni-trier.de/rec/bibtex/#1.xml}
  {dblp:#1}}
\def\mn@eprint@#1:#2:#3:#4\@nil{\def\@tempa {#1}\def\@tempb {#2}\def\@tempc
  {#3}\ifx \@tempc \@empty \let \@tempc \@tempb \let \@tempb \@tempa \fi \ifx
  \@tempb \@empty \def\@tempb {arXiv}\fi \@ifundefined
  {mn@eprint@\@tempb}{\@tempb:\@tempc}{\expandafter \expandafter \csname
  mn@eprint@\@tempb\endcsname \expandafter{\@tempc}}}

\bibitem[\protect\citeauthoryear{{Balsara}}{{Balsara}}{1995}]{Balsara1995}
{Balsara} D.~S.,  1995, \mn@doi [J. Comput. Phys.]
  {10.1016/S0021-9991(95)90221-X}, \href
  {https://ui.adsabs.harvard.edu/\#abs/1995JCoPh.121..357B} {121, 357}

\bibitem[\protect\citeauthoryear{{Benz}, {Slattery}  \& {Cameron}}{{Benz}
  et~al.}{1986}]{Benz+1986}
{Benz} W.,  {Slattery} W.~L.,   {Cameron} A.~G.~W.,  1986, \mn@doi [\icarus]
  {10.1016/0019-1035(86)90088-6}, \href
  {http://adsabs.harvard.edu/abs/1986Icar...66..515B} {66, 515}

\bibitem[\protect\citeauthoryear{{Borrow}, {Bower}, {Draper}, {Gonnet}  \&
  {Schaller}}{{Borrow} et~al.}{2018}]{Borrow+2018}
{Borrow} J.,  {Bower} R.~G.,  {Draper} P.~W.,  {Gonnet} P.,   {Schaller} M.,
  2018, Proc. 13th SPHERIC Intl. Wksh., pp 44--51

\bibitem[\protect\citeauthoryear{{{\'C}uk} \& {Stewart}}{{{\'C}uk} \&
  {Stewart}}{2012}]{Cuk+Stewart2012}
{{\'C}uk} M.,  {Stewart} S.~T.,  2012, \mn@doi [Science]
  {10.1126/science.1225542}, \href
  {http://adsabs.harvard.edu/abs/2012Sci...338.1047C} {338, 1047}

\bibitem[\protect\citeauthoryear{{Cullen} \& {Dehnen}}{{Cullen} \&
  {Dehnen}}{2010}]{Cullen+Dehnen2010}
{Cullen} L.,  {Dehnen} W.,  2010, \mn@doi [\mnras]
  {10.1111/j.1365-2966.2010.17158.x}, \href
  {http://adsabs.harvard.edu/abs/2010MNRAS.408..669C} {408, 669}

\bibitem[\protect\citeauthoryear{{Dehnen} \& {Aly}}{{Dehnen} \&
  {Aly}}{2012}]{Dehnen+Aly2012}
{Dehnen} W.,  {Aly} H.,  2012, \mn@doi [\mnras]
  {10.1111/j.1365-2966.2012.21439.x}, \href
  {https://ui.adsabs.harvard.edu/\#abs/2012MNRAS.425.1068D} {425, 1068}

\bibitem[\protect\citeauthoryear{{Deng}, {Reinhardt}, {Benitez}, {Mayer},
  {Stadel}  \& {Barr}}{{Deng} et~al.}{2019}]{Deng+2019a}
{Deng} H.,  {Reinhardt} C.,  {Benitez} F.,  {Mayer} L.,  {Stadel} J.,   {Barr}
  A.~C.,  2019, \mn@doi [\aj] {10.3847/1538-4357/aaf399}, \href
  {https://ui.adsabs.harvard.edu/abs/2019ApJ...870..127D} {870, 127}

\bibitem[\protect\citeauthoryear{{Diehl}, {Rockefeller}, {Fryer}, {Riethmiller}
   \& {Statler}}{{Diehl} et~al.}{2015}]{Diehl+2015}
{Diehl} S.,  {Rockefeller} G.,  {Fryer} C.~L.,  {Riethmiller} D.,   {Statler}
  T.~S.,  2015, \mn@doi [\pasa] {10.1017/pasa.2015.50}, \href
  {http://adsabs.harvard.edu/abs/2015PASA...32...48D} {32, e048}

\bibitem[\protect\citeauthoryear{{Genda}, {Fujita}, {Kobayashi}, {Tanaka}  \&
  {Abe}}{{Genda} et~al.}{2015}]{Genda+2015}
{Genda} H.,  {Fujita} T.,  {Kobayashi} H.,  {Tanaka} H.,   {Abe} Y.,  2015,
  \mn@doi [\icarus] {10.1016/j.icarus.2015.08.029}, \href
  {https://ui.adsabs.harvard.edu/#abs/2015Icar..262...58G} {262, 58}

\bibitem[\protect\citeauthoryear{{Gonnet}}{{Gonnet}}{2015}]{Gonnet2015}
{Gonnet} P.,  2015, SIAM J Sci. Comput., 37, C95

\bibitem[\protect\citeauthoryear{{Herant}}{{Herant}}{1994}]{Herant1994}
{Herant} M.,  1994, \memsai, \href
  {http://adsabs.harvard.edu/abs/1994MmSAI..65.1013H} {65, 1013}

\bibitem[\protect\citeauthoryear{{Hosono}, {Iwasawa}, {Tanikawa}, {Nitadori},
  {Muranushi}  \& {Makino}}{{Hosono} et~al.}{2017}]{Hosono+2017}
{Hosono} N.,  {Iwasawa} M.,  {Tanikawa} A.,  {Nitadori} K.,  {Muranushi} T.,
  {Makino} J.,  2017, \mn@doi [Publ. Astron. Soc. Jpn.] {10.1093/pasj/psw131},
  \href {https://ui.adsabs.harvard.edu/#abs/2017PASJ...69...26H} {69, 26}

\bibitem[\protect\citeauthoryear{{Hubbard} \& {MacFarlane}}{{Hubbard} \&
  {MacFarlane}}{1980}]{Hubbard+MacFarlane1980}
{Hubbard} W.~B.,  {MacFarlane} J.~J.,  1980, \mn@doi [\jgr]
  {10.1029/JB085iB01p00225}, \href
  {http://adsabs.harvard.edu/abs/1980JGR....85..225H} {85, 225}

\bibitem[\protect\citeauthoryear{{Inamdar} \& {Schlichting}}{{Inamdar} \&
  {Schlichting}}{2016}]{Inamdar+Schlichting2016}
{Inamdar} N.~K.,  {Schlichting} H.~E.,  2016, \mn@doi [\apj]
  {10.3847/2041-8205/817/2/L13}, \href
  {https://ui.adsabs.harvard.edu/#abs/2016ApJ...817L..13I} {817, L13}

\bibitem[\protect\citeauthoryear{{Kegerreis} et~al.,}{{Kegerreis}
  et~al.}{2018}]{Kegerreis+2018}
{Kegerreis} J.~A.,  et~al., 2018, \mn@doi [\apj] {10.3847/1538-4357/aac725},
  \href {https://ui.adsabs.harvard.edu/#abs/2018ApJ...861...52K} {861, 52}

\bibitem[\protect\citeauthoryear{{Kurosaki} \& {Inutsuka}}{{Kurosaki} \&
  {Inutsuka}}{2019}]{Kurosaki+Inutsuka2019}
{Kurosaki} K.,  {Inutsuka} S.-i.,  2019, \mn@doi [\aj]
  {10.3847/1538-3881/aaf165}, \href
  {https://ui.adsabs.harvard.edu/\#abs/2019AJ....157...13K} {157, 13}

\bibitem[\protect\citeauthoryear{{Lecoanet} et~al.,}{{Lecoanet}
  et~al.}{2016}]{Lecoanet+2016}
{Lecoanet} D.,  et~al., 2016, \mn@doi [\mnras] {10.1093/mnras/stv2564}, \href
  {https://ui.adsabs.harvard.edu/abs/2016MNRAS.455.4274L} {455, 4274}

\bibitem[\protect\citeauthoryear{{Leopardi}}{{Leopardi}}{2007}]{Leopardi2007}
{Leopardi} P.,  2007, PhD thesis, Sch. Math. Stat., U. New South Wales

\bibitem[\protect\citeauthoryear{{Lombardi}, {Sills}, {Rasio}  \&
  {Shapiro}}{{Lombardi} et~al.}{1999}]{Lombardi+1999}
{Lombardi} J.~C.,  {Sills} A.,  {Rasio} F.~A.,   {Shapiro} S.~L.,  1999,
  \mn@doi [J. Comput. Phys.] {10.1006/jcph.1999.6256}, \href
  {https://ui.adsabs.harvard.edu/#abs/1999JCoPh.152..687L} {152, 687}

\bibitem[\protect\citeauthoryear{{Melosh}}{{Melosh}}{2007}]{Melosh2007}
{Melosh} H.~J.,  2007, \mn@doi [Meteorit. Planet. Sci]
  {10.1111/j.1945-5100.2007.tb01009.x}, \href
  {http://adsabs.harvard.edu/abs/2007M%26PS...42.2079M} {42, 2079}

\bibitem[\protect\citeauthoryear{{Monaghan}}{{Monaghan}}{1992}]{Monaghan1992}
{Monaghan} J.~J.,  1992, \mn@doi [\araa] {10.1146/annurev.aa.30.090192.002551},
  \href {http://adsabs.harvard.edu/abs/1992ARA%26A..30..543M} {30, 543}

\bibitem[\protect\citeauthoryear{{Monaghan}}{{Monaghan}}{2012}]{Monaghan2012}
{Monaghan} J.~J.,  2012, \mn@doi [Annu. Rev. Fluid Mech.]
  {10.1146/annurev-fluid-120710-101220}, \href
  {https://ui.adsabs.harvard.edu/#abs/2012AnRFM..44..323M} {44, 323}

\bibitem[\protect\citeauthoryear{{Morris}}{{Morris}}{1996}]{Morris1996}
{Morris} J.~P.,  1996, Publ. Astron. Soc. Aust, \href
  {https://ui.adsabs.harvard.edu/#abs/1996PASA...13...97M} {13, 97}

\bibitem[\protect\citeauthoryear{{Price}}{{Price}}{2012}]{Price2012}
{Price} D.~J.,  2012, \mn@doi [J. Comput. Phys.] {10.1016/j.jcp.2010.12.011},
  \href {https://ui.adsabs.harvard.edu/\#abs/2012JCoPh.231..759P} {231, 759}

\bibitem[\protect\citeauthoryear{{Raskin} \& {Owen}}{{Raskin} \&
  {Owen}}{2016}]{Raskin+Owen2016a}
{Raskin} C.,  {Owen} J.~M.,  2016, \mn@doi [\apj]
  {10.3847/0004-637X/820/2/102}, \href
  {http://adsabs.harvard.edu/abs/2016ApJ...820..102R} {820, 102}

\bibitem[\protect\citeauthoryear{{Reinhardt} \& {Stadel}}{{Reinhardt} \&
  {Stadel}}{2017}]{Reinhardt+Stadel2017}
{Reinhardt} C.,  {Stadel} J.,  2017, \mn@doi [\mnras] {10.1093/mnras/stx322},
  \href {https://ui.adsabs.harvard.edu/\#abs/2017MNRAS.467.4252R} {467, 4252}

\bibitem[\protect\citeauthoryear{{Saff} \& {Kuijlaars}}{{Saff} \&
  {Kuijlaars}}{1997}]{Saff+Kuijlaars1997}
{Saff} E.~B.,  {Kuijlaars} A. B.~J.,  1997, The Math. Int., 19, 5

\bibitem[\protect\citeauthoryear{{Schaller}, {Gonnet}, {Chalk}  \&
  {Draper}}{{Schaller} et~al.}{2016}]{Schaller+2016}
{Schaller} M.,  {Gonnet} P.,  {Chalk} A. B.~G.,   {Draper} P.~W.,  2016,
  \mn@doi [Proc. PASC 16 Conf.] {10.1145/2929908.2929916}, pp 2:1--2:10

\bibitem[\protect\citeauthoryear{{Schaller et al.}}{{Schaller et
  al.}}{2018}]{Schaller+2018}
{Schaller et al.} M.,  2018, {SWIFT: SPH With Inter-dependent Fine-grained
  Tasking}, Astrophysics Source Code Library (\mn@eprint {ascl} {1805.020})

\bibitem[\protect\citeauthoryear{{Slattery}, Benz  \& Cameron}{{Slattery}
  et~al.}{1992}]{Slattery+1992}
{Slattery} W.~L.,  Benz W.,   Cameron A.~G.~W.,  1992, \mn@doi [\icarus]
  {10.1016/0019-1035(92)90180-F}, \href
  {http://adsabs.harvard.edu/abs/1992Icar...99..167S} {99, 167}

\bibitem[\protect\citeauthoryear{{Springel}}{{Springel}}{2005}]{Springel2005}
{Springel} V.,  2005, \mn@doi [\mnras] {10.1111/j.1365-2966.2005.09655.x},
  \href {http://adsabs.harvard.edu/abs/2005MNRAS.364.1105S} {364, 1105}

\bibitem[\protect\citeauthoryear{{Springel}}{{Springel}}{2010}]{Springel2010}
{Springel} V.,  2010, \mn@doi [Annu. Rev. Astron. Astrophys.]
  {10.1146/annurev-astro-081309-130914}, \href
  {https://ui.adsabs.harvard.edu/#abs/2010ARA&A..48..391S} {48, 391}

\bibitem[\protect\citeauthoryear{Tillotson}{Tillotson}{1962}]{Tillotson1962}
Tillotson J.~H.,  1962, General Atomic Report, GA-3216, 141

\bibitem[\protect\citeauthoryear{Wallace, Sidles  \& Danielson}{Wallace
  et~al.}{1960}]{Wallace+1960}
Wallace D.~C.,  Sidles P.~H.,   Danielson G.~C.,  1960, \mn@doi [J. Appl.
  Phys.] {10.1063/1.1735393}, 31, 168

\bibitem[\protect\citeauthoryear{{Wang} \& {White}}{{Wang} \&
  {White}}{2007}]{Wang+White2007}
{Wang} J.,  {White} S. D.~M.,  2007, \mn@doi [\mnras]
  {10.1111/j.1365-2966.2007.12053.x}, \href
  {https://ui.adsabs.harvard.edu/#abs/2007MNRAS.380...93W} {380, 93}

\bibitem[\protect\citeauthoryear{Waples \& Waples}{Waples \&
  Waples}{2004}]{Waples+Waples2004}
Waples D.~W.,  Waples J.~S.,  2004, \mn@doi [Nat. Resour. Res.]
  {10.1023/B:NARR.0000032647.41046.e7}, 13, 97

\bibitem[\protect\citeauthoryear{{Willis}, {Schaller}, {Gonnet}, {Bower}  \&
  {Draper}}{{Willis} et~al.}{2018}]{Willis+2018}
{Willis} J.~S.,  {Schaller} M.,  {Gonnet} P.,  {Bower} R.~G.,   {Draper} P.~W.,
   2018, \mn@doi [Adv. Parallel Comp.] {10.3233/978-1-61499-843-3-507}, 32, 507

\bibitem[\protect\citeauthoryear{{Woolfson}}{{Woolfson}}{2007}]{Woolfson2007}
{Woolfson} M.~M.,  2007, \mn@doi [\mnras] {10.1111/j.1365-2966.2007.11498.x},
  \href {http://adsabs.harvard.edu/abs/2007MNRAS.376.1173W} {376, 1173}

\makeatother
\end{thebibliography}




\appendix

\section{Planetary Profiles}
\label{sec:profiles}
This section details the creation of radial profiles for the model planets.
The main inputs for a profile are the total mass,
the number of layers and their materials,
the surface pressure and temperature,
and estimates for the outer radius and any internal boundary radii
that we will later refine.
To set each layer's material,
we must define the equation of state (EoS),
a conversion between temperature and internal energy
e.g. the specific heat capacity and cold curve,
and an expression for how heat is transferred
e.g. isothermal or adiabatic.

We iterate inwards in thin spherical shells
from the surface to the centre --
not to be confused with the much thicker shells
we define in \S\ref{sec:method:shells}
to arrange simulation particles in the resulting sphere.
The density at the surface is first found using the EoS
with the input pressure and temperature.
Assuming a constant density within this very thin shell,
the mass of the shell is calculated to find
the pressure at the inner shell boundary
that would be required for hydrostatic equilibrium.
The density and temperature that provide
this pressure at the inner shell boundary
are then found using the EoS and the heat transfer ($\rho$--$T$) relation.
This process is repeated for the next shell until reaching the centre.

The temperature and pressure are continuous
across any internal layer boundaries,
so this iteration continues into the core,
until the input total mass has been used up.
If the input radii for the outer surface and any inner boundaries are accurate,
then the central shell should use up the final available mass
just as its inner boundary reaches the centre.
However, if any of these input radii are too large or too small,
then either the mass will be used up before reaching the centre
or the centre will be reached with some mass still remaining.
In this case, we modify the input radii and repeat the process,
until the mass discrepancy is a tiny fraction of the total mass.

For our test model of a simple Earth-mass planet 
in \S\ref{sec:results:particles},
the inputs were the Earth's mass,
the Tillotson granite EoS \citep{Tillotson1962,Melosh2007},
and an isothermal temperature of 300~K,
leading to an outer radius of 1.036~$R_\oplus$.
We chose a constant specific heat capacity of
710~J~K$^{-1}$~kg$^{-1}$ \citep{Wallace+1960,Waples+Waples2004}.

The resulting density (and temperature or internal energy) profile
can then be used to create a set of particle initial conditions,
as described in \S\ref{sec:methods:particles}.
This approach is the same for more complicated planets with multiple layers
and discontinuities in material and density,
such as the proto-Uranus and impactor used in \S\ref{sec:results:uranus}
with full details in \citet{Kegerreis+2018}.

\section{Tillotson Sound Speed}
\label{sec:tillotson}

In addition to the pressure, density, and thermal properties of a material,
the EoS is also important for determining the sound speed.
In smoothed particle hydrodynamics (SPH),
the sound speed is used both to control the simulation timestep --
to ensure that sound waves do not travel
further than the distance between neighbouring particles in one step --
and as part of the artificial viscosity calculation
that controls the behaviour of shocks \citep{Price2012}.

The popular Tillotson EoS does not include an expression for the sound speed,
$c$, but it can be derived from
the partial derivative of the pressure, $P$,
with respect to the density, $\rho$, at constant entropy, $S$:
\begin{equation}
  c^2 = \left. \dfrac{\partial P}{\partial \rho} \right|_S \;,
  \label{eqn:soundspeed}
\end{equation}
which we can calculate from Tillotson's
$P$, $\rho$, and specific internal energy $u$, using 
${\rm d}u = T {\rm d}S - P {\rm d}V = T {\rm d}S + (P/\rho^2) {\rm d}\rho $.

The Tillotson pressure is separated into a condensed or cold state
and an expanded and hot state \citep{Tillotson1962}.
Using the standard definitions of $\eta \equiv \rho / \rho_0$,
$\mu \equiv \eta - 1$, $\nu \equiv 1/\eta - 1$,
and $\omega \equiv u / (u_0 \eta^2) + 1$,
these two pressure formulae are
\begin{align}
  P_{\rm c} &= \left(a + \dfrac{b}{\omega}\right) \rho u
  	+ A\mu + B\mu^2 \\
  P_{\rm e} &= a \rho u + \left(\dfrac{b \rho u}{\omega}
  	+ A \mu e^{-\beta \nu}\right) e^{-\alpha \nu^2} \;,
\end{align}
where $\rho_0$, $a$, $b$, $A$, $B$, $\alpha$, $\beta$, $u_0$, $u_{\rm{iv}}$,
and $u_{\rm{cv}}$
are material-specific parameters for the EoS \citep{Melosh2007}.
In the hybrid state, the pressure is a linear combination of the two:
\begin{equation}
  P_{\rm h} = \dfrac{(u - u_{\rm{iv}}) \,P_{\rm e} + (u_{\rm{cv}} - u)
        \,P_{\rm c}}{u_{\rm{cv}} - u_{\rm{iv}}} \;.
\end{equation}
For SWIFT, the minimum pressure is set to 0.

Using Eqn.~\ref{eqn:soundspeed}, the sound speeds for each state are
\begin{align}
  c_{\rm c}^2 = &\;
  	\dfrac{P_{\rm c}}{\rho} \left[1 + a + \dfrac{b}{\omega}\right]
  	+
    \dfrac{b (\omega-1)}{\omega^2} \left[2 u - \dfrac{P_{\rm c}}{\rho}\right]
  \nonumber
  \\&
    +
    \dfrac{1}{\rho}\left[A + B\left(\eta^2 - 1\right)\right]
  \\
  c_{\rm e}^2 = &\;
  \dfrac{P_{\rm e}}{\rho} \left[1 + a + \dfrac{b}{\omega} e^{-\alpha \nu^2}\right]
  +
  \left\{\dfrac{b\rho u}{\omega^2 \eta^2}
    \left[\dfrac{1}{u_0 \rho}\left(2u - \dfrac{P_{\rm e}}{\rho}\right)
      + \dfrac{2\alpha\nu\omega}{\rho_0} \right]\right.
  \nonumber
  \\&
  +
  \left. \dfrac{A}{\rho_0}
    \left[1 + \dfrac{\mu}{\eta^2} \left(\beta + 2\alpha\nu - \eta\right)
    \right] e^{-\beta\nu} \right\} e^{-\alpha\nu^2} \;,
\end{align}
and the hybrid state is the equivalent linear combination:
\begin{equation}
  c_{\rm h}^2 = \dfrac{(u - u_{\rm{iv}}) \,c_{\rm e}^2 + (u_{\rm{cv}} - u)
        \,c_{\rm c}^2}{u_{\rm{cv}} - u_{\rm{iv}}} \;.
\end{equation}
For SWIFT, a minimum sound speed is set
using the uncompressed density and bulk modulus: $\sqrt{A/\rho_0}$.

\citet{Reinhardt+Stadel2017} did this same calculation
(with slightly different notation),
but their $c_{\rm c}^2$ has a typo $A$ instead of $a$ in the first term
and their $c_{\rm e}^2$ has swapped the sign of
$\left(2u - P_{\rm e}/\rho\right)$,
which would change the sound speed by $\sim$10\%.


\bsp	
\label{lastpage}
\end{document}